\documentclass[11pt]{article}
\usepackage{braket}
\usepackage{graphicx}
\usepackage[margin=1in]{geometry}

\usepackage{makeidx}  
\usepackage{ifpdf}
\usepackage{url}

\title{Elementary Cellular Automata as Non-Cryptographic Hash Functions}
\date{June 2025}
\author{Daniel McKinley}

\begin{document}

\maketitle

\section{Introduction}

Ten of the 256 elementary cellular automata (ECA) rules are explored as a non-cryptographic hash function using a lossy compression error-minimization function that operates on input data in 4x4 binary cells \cite{Wolfram}. The hash's key properties are that the codewords are unique and evenly distributed, has a lossy inverse, hashed data can be operated on efficiently within a hashed state, and through bitmap file processing shows a clear application to edge detection at shallow depths of iteration.  The spacing of the individual tile neighborhoods parallel the nested $2^n$ structure of the Fast Fourier Transform (FFT) and Fast Walsh-Hadamard Transform. General algorithm outline, specific ECA rules, and aggregate properties are discussed. It is implemented in Java and more images and gifs and code links are available at the website\cite{dmwebsite}.\\

\subsection{Hash Function Background}

Hash functions can be described as indexing functions analogous to alphabetical order. If I give you a random word you've never heard before you can still look it up in the dictionary because they are in alphabetical order. It would be more difficult to look it up in a book ordered by numbers of letters because of conflicts when different words have the same number of letters, and this is the difference between different kinds of hashes. An extension of this idea can do the same thing for any kind of data including audio and visual. Databases use hash functions to index large amounts of data and web servers use hashes to distribute workloads. If a site gets a billion visitors a day the work has to be divided up between sub servers by hashing certain reference data such as the client's id data. Databases keep frequently used data closer to the top of the stack, using hashes and bloom filters to determinine whether requested data is already up front in the cache or that a deeper database has to be called, saving significant amounts of time and space.\\

There are two kinds of hashes called cryptographic hashes and non-cryptographic hashes. Cryptographic hashes are used in password storage, encryption, blockchain currency, and other secure applications. A survey of literature using automata as a hash function show a focus on the cryptographic kind. \cite{app14219719} \cite{article} \cite{Rajeshwaran2019CellularAB} The hashes described here are almost the opposite of cryptographic because of the inverses, retroactive hashing, minimal avalanche, and linear operation component rules.\\

\subsection{Elementary Cellular Automata Background}

Elementary Cellular Automata are a 1 dimensional operation with a set of 256 8 bit binary truth tables numbered 0..255 called Wolfram codes \cite{Wolfram}. They do this 1D operation in parallel on a binary string, taking in 3 bits at each location, itself and its two neighbors on either side, binary summing these 3 input bits to a value 0..7, and looking up that value's bit in the Wolfram code as the output. Doing this operation repeatedly produces behavior falling into four classes of behavior: uniform, periodic/stable, chaotic, and complex. This hash project sieves through all 256 elementary rules, but uses a small square wrapped output space rather than going to infinity on either side.

\section{Main Algorithm}
This hash algorithm is a kind of lossy compression that operates on 4x4 wrapped cells of binary input data that minimizes errors in compression on a small scale looped over all the input data, and the same algorithm works for maximizing error. Within each cell, row 0 is the input neighborhood and rows 1, 2, and 3 are the ECA rule's output. All 16 possible row 0 inputs are calculated for a given input and then scored so that each bitwise discrepancy between the codeword's output and the input is summed with a weight of $2^{row}$ or $2^{column}$. The input neighborhoods that minimize and maximize the error score are noted as the codeword pair for that 4x4 input and each codeword is 4 bits. Doing this procedure for all  $2^{16}=65536$ possible 4x4 input neighborhoods produces a truth table for a particular ECA rule and minMax rowColumn parameters.\\

\begin{center}
\includegraphics{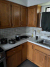}\\
Original Image\\
\end{center}
\begin{center}
\includegraphics{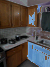}\\
Frame 1\\
\end{center}
\begin{center}
\includegraphics{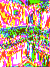}\\
Frame 2\\
\end{center}
\newpage

\begin{center}
There are $2^{16}=65536$ binary 4x4 arrays, wrapped by column\\
$2^4$ possible $row_0$ neighborhoods for a given ECA rule\\
The ECAfunction(input) is output for rows 1, 2, 3
These 16 outputs are scored for errors by\\
Summing the discrepancies between originalInput and codewordOutput\\
Weighted by either $2^{row}$ or $2^{column}$\\
\[  \sum_{r=0}^{3} \sum_{c=0}^{3} 2^r ( compressionAttempt_{r c} \oplus original_{r c}) \]\\
 The minimizing and maximizing values of all possible input are noted\\
 as the codeword pair of the original binary matrix for a given ECA rule\\
 \end{center}
 \begin{center}
\includegraphics{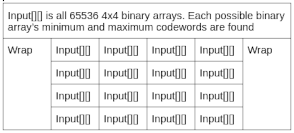}
\includegraphics{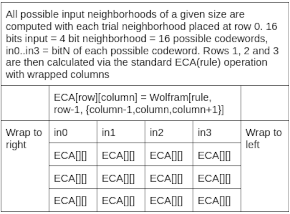}
\includegraphics{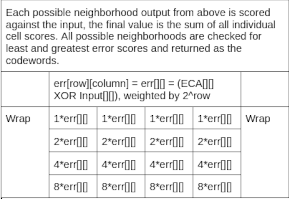}
\end{center}

Having generated the truth table for a given ECA rule, it is applied to a 2D binary array and bitmap image as follows and with a slight variation is done for 1D input. For every (wrapped) (row,column) location in the input, the local 4x4 cell hash input is the sum of it and its neighbors $2^d$ away, $(row,column)..((row+2^d),(column+2^d))$ where d is depth of iteration. The location's value is replace with the respective minimizing or maximizing codewords that best represent its neighborhood by minimizing error. The comparing of neighbors of powers of 2 distance away is the same stucture as the FFT and Fast Walsh Transforms with hashes instead of sums. When the iteration's neighbor distance is $log2_{row}$ or $log_2(column)$, whichever is greater, every bit has influenced every other bit and the avalanche property can be analyzed. Since every neighborhood has a unique solution, any size of any given input has a unique hash if hashing in place and not compressing.\\

\begin{center}
Below is a Fast Walsh-AlgorithmCode.Hadamard Example \cite{enwiki:1261916659}\\
\includegraphics{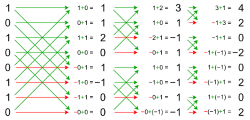}\\

Doing the above with the powers of 2 in reverse order is the following\\
1 dim version; instead of a sum term it's a hash; 2D is two axes\\ 
Find the codewords of the codewords, twice as far apart each time\\
\includegraphics{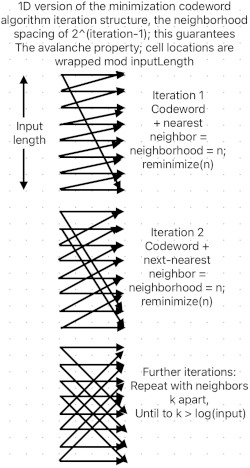}\\
\end{center}

\section{Hash properties}

\subsection{Two sets of Eight Rules}
Within the 256 ECA rules, there are 8 $[0,15,51,85,170,204,240,255]$ using a row weighted errorScore that have the properties of unique codewords for any given input and every codeword occurs the same number of times with relatively even distribution of codewords across the 65536 neighborhoods in the truth table.  Since every lookup is unique at small scale, every general input has a unique solution and there are no small or large scale collisions. Changing a single bit in the input data may not change a single codeword's output overall but by using the entire set including the columnError collisions can be prevented entirely. In principle you need the columnError set to guarantee no collisions but in practice even a single bit change showed up in the loops tracking two very similar bitmaps through the hash process.\\

There is another overlapping subset of 8, $[0,15,85,90,165,170,240,255]$ whose errorScore weight is $2^{column}$ rather than $2^{row}$. Again, the codewords are distributed perfectly evenly, there are unique solutions for every input, and has a minimizing and a maximizing codeword set. These lists share a few members with the XOR-additive list. \cite{xorAdditive}. Most properties apply to both with a few exceptions.\\

0 and 255 are included because of the 4 rows of the output matrix, 1 is neighborhood input and 3 are output, and so still produce an errorScore and a unique solution. In the hashLogicOpTransform class while 255 minimizations have a valid transform in its truth table, it has to be manually overidden to zero, adding a constant to the O(N). All other codewords with valid operations work seamlessly. Another reason they're not completely trivial is that a complete codeword set can be encoded as trigintaduonions, hypercomplex rings with degree 5, two degrees past octonions, having 32 elements, with no negative members when using 4 bit codewords because the sixteenth's place is the negative bit. Having a unique distinct solution, the trigintaduonions lend themselves easily to physical interpretations of hash sets, can be an orthonormal basis for data structures, and allows for tight integration with GF($2^m$).  \\

\subsection{Retroactive Hashing}
Hashed input can be operated on within the row weighted rule set without the original input and without inverting it. These operations only partially apply to the column weighted set, some of the rule-logic pairs have a uniform operation transform and some do not. Take any 2 of 16 codeword-generated 4x4 cells, logicGate(A) them together and rehash, the result is the same as the original two codewords logicGate(B) together for all values in the truth table. This shift of logic operations within a hash is uniform within an ECA rule hash and extends to any depth of iteration; the hash algorithm transforms not only the input data but also the relative logic gates within codewords. \\

For example, you can retroactively apply a watermark to a hashed image or IP address bitmask without the original input and without lossy and expensive inversion. Hash the bitmask to the same depth as the original and lookup the appropriate logic gate operation between the hashed image and the hashed bitmask. These equivalences appear to carry over to size 8 in random testing. Two pairs of columnError gates have only 3=A, 5=B, 10=!A, and 12=!B, which allow for locally customizable reads and writes as a kind of targetable selector switch, but are not universal. Rules 90 and 165 have an XOR and XNOR gate, which are also not universal.  \\

This happens through a transformation of logic operation because of the linear nature of the rules used. Rules 15, 51, and 85 simply pass the left middle or right input bit with no other operation or dependency, 170, 204, and 240 are the complements, and 90 and 165 are the parity of the left and right bits and so define properties of the columnWeighted set. Active development on the project includes workarounds for the missing operations. Experimental testing has verified the rowError minimizing set on odd numbered iterations, with promising results for others using several different initial bitmap raster breakdown loops. Some of these have zero errors and some every single bit is negated; in experimentation there appears to be  significance to the parity of the iteration relative to the type of RGB code and hexadecimal vs binary, and that there is a gate negation pattern where complements have to be taken at certain points relative to their position within the codeword set.\\
\newpage
\begin{center}
Rows are logic gates, AND = 8, OR = 14, XOR = 6...\\
Columns are the minMax/rowColumn codewords\\
\hfill \break
Row weighted error\\
$[0,15,51,85,170,204,240,255,0,15,51,85,170,204,240,255]$ 8 min and 8 max\\
Every column has one of every gate\\
\hfill \break

00 15 15 15 00 00 00 00 15 00 00 00 15 15 15 15 \\
01 07 07 07 01 01 01 01 14 08 08 08 14 14 14 14 \\
02 11 11 11 02 02 02 02 13 04 04 04 13 13 13 13 \\
03 03 03 03 03 03 03 03 12 12 12 12 12 12 12 12 \\
04 13 13 13 04 04 04 04 11 02 02 02 11 11 11 11 \\
05 05 05 05 05 05 05 05 10 10 10 10 10 10 10 10 \\
06 09 09 09 06 06 06 06 09 06 06 06 09 09 09 09 \\
07 01 01 01 07 07 07 07 08 14 14 14 08 08 08 08 \\
08 14 14 14 08 08 08 08 07 01 01 01 07 07 07 07 \\
09 06 06 06 09 09 09 09 06 09 09 09 06 06 06 06 \\
10 10 10 10 10 10 10 10 05 05 05 05 05 05 05 05 \\
11 02 02 02 11 11 11 11 04 13 13 13 04 04 04 04 \\
12 12 12 12 12 12 12 12 03 03 03 03 03 03 03 03 \\
13 04 04 04 13 13 13 13 02 11 11 11 02 02 02 02 \\
14 08 08 08 14 14 14 14 01 07 07 07 01 01 01 01 \\
15 00 00 00 15 15 15 15 00 15 15 15 00 00 00 00 \\
\hfill \break
Column weighted error\\
$[0,15,51,90,165,204,240,255,0,15,51,90,165,204,240,255]$ 8 min and 8 max\\
77 represents no operation valid on the codeword-gate\\ 
\hfill \break

00 77 77 00 77 00 00 00 15 77 77 77 15 15 15 15 \\
01 77 77 77 77 01 01 01 14 77 77 77 77 14 14 14 \\
02 77 77 77 77 02 02 02 13 77 77 77 77 13 13 13 \\
03 03 03 77 77 03 03 03 12 12 12 12 12 12 12 12 \\
04 77 77 77 77 04 04 04 11 77 77 77 77 11 11 11 \\
05 05 05 77 77 05 05 05 10 10 10 10 10 10 10 10 \\
06 77 77 06 77 06 06 06 09 77 77 77 09 09 09 09 \\
07 77 77 77 77 07 07 07 08 77 77 77 77 08 08 08 \\
08 77 77 77 77 08 08 08 07 77 77 77 77 07 07 07 \\
09 77 77 77 09 09 09 09 06 77 77 06 77 06 06 06 \\
10 10 10 10 10 10 10 10 05 05 05 77 77 05 05 05 \\
11 77 77 77 77 11 11 11 04 77 77 77 77 04 04 04 \\
12 12 12 12 12 12 12 12 03 03 03 77 77 03 03 03 \\
13 77 77 77 77 13 13 13 02 77 77 77 77 02 02 02 \\
14 77 77 77 77 14 14 14 01 77 77 77 77 01 01 01 \\
15 77 77 77 15 15 15 15 00 77 77 00 77 00 00 00 \\
\end{center}
\newpage
\subsection{Inverse}

The inverse uses a voting system where the codeword neighborhoods add or subtract the appropriate error weight at every location in their zone of influence. Each codeword can be inverted separately or as a complete set. The codeword is the algorithm's best guess at what the input was, so the inverse is distributing it's codeword neighborhood's errorScore weighted best guess at the pixels they're relevant to, and are tallied and accounted for as output 0s and 1s. A codeword votes its lossily reproduced neighborhood with the proper string of positives and negatives  for value 0..1, minimization..maximization, and rowError..columnError. The inverse is deterministic, any given hash has only one inverse.\\

Experimentation with inverting hashed 2 byte and 4 byte RGB code bitmaps using various initial raster processing loops and hexadecimal vs binary parameters resulted in a wide variation of error rates, from 0.7 errors/bit to 0.02 errors/bit. Some versions show great consistency across codewords with error rates roughly the same as the error rate of their respective truth tables, about 3/8 errors/bit. Other versions showed wide variations in error rates across codewords; some versions of the inverse did better depending on the parity of the iteration as compared to number of bytes in the RGB code. Sometimes an inverse would recreate the original frame with a low error rate and others quickly drop to noise. There may be a perfect inverse for some subset of codewords, it is an active area of development.\\

\subsection{Avalanche and Collisions}
This algorithm does not display the avalanche property; the threshold for testing is when every pixel's RGB code has had an opportunity to influence every other pixel, or $log_2(imageWidth)$ or $log_2(imageHeight)$, whichever is greater. Experimental results with bitmaps show that even a 1 bit change alters at least one codeword set solution and that this change propagates forward with discrepancies tracking through the process like trees. The truth tables of any single codeword for any input is not unique, however if checking the entire set of 32 codewords at once the set is unique. Since the set of truth tables is distinct, any given input of any size or shape is unique and collisions don't happen for a given size. Tracking the propagation shows that at least one codeword in the set is changed with any bitmap and that it branches from there with every iteration. It could be described as a minimal avalanche instead of maximal avalanche, just enough to prove uniqueness and propagation of small initial changes.\\

\subsection{Input Sizes}
This algorithm can be used to make any sized hash output. It's implemented to either hash-in-place so that the data stays the same size or as compression with the data size quartering each iteration. To implement any size output, first hash to the avalanche point as the fixed data size version, then as compression to the size desired and/or pad with zeroes at the beginning or the end. Rather than padding with zeroes one can use an inversion to expand the hash to a larger size, however inversion is more computationally expensive than compression. Another option is to hash to the avalanche point and take a subset of it. If using every minMax row-column codeword set, hash to desired size and then hash the hashes.\\
\subsection{Edge Detection}

Visual inspection of hashed bitmaps show a clear application to edge detection. Natural parts of images such as trees and shrubs appear chaotic while man made surfaces such as countertops, walls, and garage doors merge to a single color that tends to change sharply over edges and corners.\\

\subsection{Codeword Symmetries}
The 32 codewords for any 16 bit input display some symmetries generated by the same algorithm as the left-right-black-white symmetries of the ECA Wolfram code symmetry groups, applied to 4 bits instead of 8. The left right symmetry is reversing the bit's place order and the black white symmetry is done by reversing the truth table and taking the complement and the left right black white is doing both operations. Applying this to the 4 bit codewords yields some symmetries between rules but no two codeword sets have the same so there are no similar groups to the 88 independent sets in the ECA.\cite{Wolfram}\\

\section{Project Overview}

\subsection{Implementation and testing}

The algorithm is prototyped on 2-byte RGB *.bmp bitmap files and some 4-byte, with most photos taken on an iPhone and converted with GIMP. Animated *.gif files and more images are available at my website. You can see the areas of the image hashing 2 by 2 and slowly dissolving into avalanche territory that eventually just looks like noise everywhere. The current image was chosen because you can clearly see parts of the image doubling itself as the avalanche property slowly takes over. The code works with any size bitmap or any integer array in general. \\

There are several tests avalailable to verify the integrity of the algorithm. A primary indicator is that every rowError codeword has a single internal hash logic gate solution across all 16x16 elements in the truth table for every 0..65535 value in its truth table, though this is less obvious with the column weighted set because it is incomplete. The set of images and gifs produced visually verify that the algorithm is processing as predicted. Loops that check for uniqueness and even distribution are easily verified in the 4x4 size. The numbers of 1 bits/frame are tracked to verify non empty non static arrays throughout the process. Areas of active development include the search for better inverses and workarounds to the incomplete columnError retroactive hash logic transform.\\

The O(N) of the hash is highly dependent on the size of the input, number of bits per pixel, and level of raster processing whether hexadecimal or single bit. Once the codeword truth tables are calculated, an individual tile becomes a small sum and constant term array lookup, however this is multiplied times the area of the input, possibly each of every 32 codewords, and possibly times $\log_2 sizeInput$ to get to the avalanche point . The inverses are more expensive per iteration, since the sum and lookup becomes a generate and iterate.\\

\newpage

\subsection{Other rules, shapes, and sizes}

The codeword sets can be generated with any size array, with only square size 4 being fully tested and a square 8 showing the same properties in random testing. Within this prototype project, calculation of truth table lengths of $2^{16}$ using all 4x4 binary grids are acceptable, lengths of $2^{64}$ using 8x8 grids would be challenging at this point. At size 8 there are $2^8=256$ possible codewords, which means that while you can't calculate the whole truth table at once you can calculate the codewords for any input individually on the fly. The codeword set's uniqueness and distribution properties seem to apply at size 8 by testing random codeword input tiles; it is only exhaustively tested for size 4. The internal hash logic transforms can also still be easily calculated for size 8 because you only have to test codewords rather than the entire set of inputs. Future iterations of the project can implement tile rectangles, allowing for easier 3D hashing. \\

Out of the other 0-255 ECA rules, some do better than these particular 8 at lossy compression, losing only 3/16 bits instead of 6/16 bits with most of these 10. In particular the parity rules 90, 165, 102, 153, 105, 150 rank near the top, connected to several of these 10 via the property of XOR-additiveness \cite{xorAdditive}. However none outside of these 10 have unique solutions or even distributions in either row or column weighted, maxxed or minned.\\

\bibliographystyle{plain}
\bibliography{ECAhashPaperDM.bib}

\end{document}